\documentclass[10pt,twoside]{article}
\oddsidemargin = \evensidemargin
\usepackage{latexsym}
\usepackage{amssymb}
\usepackage{amsmath}
\usepackage{epsfig}
\usepackage{url}

\begin{document}

{\small \noindent \copyright \ 2019 Imprint Academic \hfill Mind \& Matter Vol.~17(1), pp.~95--121}
\vspace{0.7cm}

{
\Large
\centerline{\bf How Dualists Should (Not) Respond to}
\centerline{\bf the Objection from Energy Conservation}
}

\bigskip

\centerline{\sl Alin C.~Cucu}
\centerline{\sl International Academy of Philosophy}
\centerline{\sl Mauren, Liechtenstein}

\smallskip
\centerline{and}

\smallskip
\centerline{\sl J.~Brian Pitts}
\centerline{\sl Faculty of Philosophy and Trinity College}
\centerline{\sl University of Cambridge, United Kingdom}

\setcounter{page}{95}
\thispagestyle{empty}

\renewcommand{\textheight}{18.5cm}

\begin{abstract}
\smallskip

The principle of energy conservation is widely taken to be a serious difficulty for interactionist dualism (whether property or substance). Interactionists often have therefore tried to make it satisfy energy conservation. This paper examines several such attempts, especially including E.~J.~Lowe's varying constants proposal, showing how they all miss their goal due to lack of engagement with the physico-mathematical roots of energy conservation physics: the first Noether theorem (that symmetries imply conservation laws), its converse (that conservation laws imply symmetries), and the locality of continuum/field physics. Thus the ``conditionality response'', which sees conservation as (bi)conditional upon symmetries and simply accepts energy non-conservation as an aspect of interactionist dualism, is seen to be, perhaps surprisingly, the one most in accord with contemporary physics (apart from quantum mechanics) by not conflicting with mathematical theorems basic to physics. A decent objection to interactionism should be {\it a posteriori}, based on empirically studying the brain. 
\end{abstract}
\bigskip
\bigskip

\noindent {\large\bf 1. The Objection from Energy Conservation}

\vspace{0.1cm}
 \hspace*{0.1 cm}{\large\bf  Formulated}

\medskip
Among philosophers of mind and metaphysicians, it is widely believed that the principle of energy conservation poses a serious problem to interactionist dualism. Insofar as this objection afflicts interactionist dualisms, it applies to property dualism of that sort (i.e., not epiphenomenalist) as much as to substance dualism (Crane 2001, pp.~40, 43, 50).\footnote{The term ``substance dualism'' suggests willingness to call the body a substance, which one might deny even while accepting strong claims about the independence of the mind from the body.}  Though less prominent than energy conservation, momentum conservation is, as a physical principle, on equal footing with energy conservation and poses the same kind of objection (Cornman 1978). In fact, Leibniz's conservation-based objection against substance dualism appealed to the conservation of momentum as well as energy ({\it vis viva}) (Garber 1983a, Leibniz 1985, p.~156). Any successful treatment must address the conservation of energy and momentum (as well as any other relevant laws that might seem violated, such as the conservation of angular momentum), a fact that thwarts some naive attempts to make interactionist dualism physics-friendly. 

If there were a good objection from energy conservation, then presumably philosophers of physics would endorse it on such occasions as they treat the philosophy of mind (which are admittedly rare, in contrast to the logical positivists' discussions). While Bunge (1980, p.~17) endorses the objection from energy conservation (albeit without any clear argument; Pitts 2019a), Butterfield (1997, p.~142) explicitly rejects it: 
\begin{quote} \small
... [A] traditional argument against interactionism is flawed, because of this false picture 	of physics. ...  The idea is that any causal interaction between mind and matter would 	violate the principle of the conservation of energy. ...  (Though traditional, the argument 	is still current; for example, Dennett endorses it (1991, pp. 34--35).) 

This argument is flawed, for two reasons. The first reason is obvious: who knows how 	small, or in some other way hard to measure, these energy gains or losses in brains might 	be? ... But the 	second reason is more interesting, and returns us to the danger of 	assuming that physics is cumulative. Namely, the principle of the conservation of energy 	is not sacrosanct [having been questioned in the past and present in quantum contexts]. 

In short:  physicalists need to be wary of bad reasons to think physicalism is true, arising 	from naivety about physics.
\end{quote}
One of us has recently explained in further detail some of what Butterfield presumably had in mind (Pitts 2019a). 

 \renewcommand{\topmargin}{1.2cm}
Interestingly, one does not always see a proper formulation of the objection from energy conservation in the literature. At times worries about energy conservation are conflated with some version of the ``causal nexus problem'' (see, e.g., Dennett 1991, pp.~34f, Fodor 1998, McGinn 1999, p.~92, Westphal 2016, pp.~41-44). The causal nexus problem involves the intuition that there does not seem to be any causal interface between non-physical and physical entities that would allow the non-physical entities to interact with the physical world. In correspondence with Descartes, Gassendi and Princess Elisabeth (Garber 1983b) expressed difficulty seeing how a non-spatial soul could interact with a body that has the property of extension; the worry and related ones continue to this day. The problem is also aggravated by Descartes's making souls non-spatial, a view that not all early modern or contemporary dualists share. The metaphysical type of objection, though significant (but see Ducasse 1951, Chap.~18, for a dissenting view), is quite distinct from Leibniz's later physical objection from conservation laws. 

A version of the objection from energy conservation that addresses the conservation issue without confusing it with more general causal worries can be found from John Searle (2004, p.~42):
\begin{quote} \small
Physics says that the amount of matter/energy in the universe is constant, but substance dualism seems to imply that there is another kind of energy, mental energy or spiritual energy, that is not fixed by physics. So if substance dualism is true then it seems that one of the most fundamental laws of physics, the law of conservation, must be false. 
\end{quote}
Searle's version is interesting in two respects. First, his version of the principle of energy conservation -- which differs substantially from the one modern physics holds (on which more below) -- is probably the one most widely used by non-physicists. Second, he aptly points out that the crux lies in an apparent contradiction between dualism and the principle of energy conservation, assuming that in this case dualism will have to yield, given the fundamentality of the principle of energy conservation. 

His talk of mental energy, by contrast, is confused. We hardly have a notion what ``mental energy'' might be; the better one understands physical energy in terms of the relevant mathematics (Noether 1918, Kosmann-Schwarzbach 2011),
the less sense the idea of mental energy makes (see Sec.~5 for more on this). Furthermore, mental energy would only be relevant in this context if it affected the physical world, which influence Searle neglects to mention explicitly in this passage. (If the archangel Michael had some kind of mental energy but it did not affect the physical world, his doings would be irrelevant to interactionist dualism and would not violate any physical conservation law.)  What Searle should say is that if nonphysical minds\footnote{There are dualistic accounts which construe the mind as partly physical (e.g., Collins 2011b).} affected the physical world (interactionist dualism), then they would cause energy changes in the physical world, resulting in a violation of the principle of energy conservation, but since that principle is too fundamental to be false, such interactionist dualism must be false. A bit more formally, the argument could run as follows as a formalized objection from energy conservation:

\begin{itemize}

\vspace{-0.15cm}\item[P1] If nonphysical minds influenced the physical universe, then the energy of the physical universe would not be constant ({\it ex hypothesi}).

\vspace{-0.15cm}
\item[P2] The energy of the physical universe is constant (with physical necessity).

\vspace{-0.15cm}
\item [C] Therefore, it is false that nonphysical minds influence the physical universe (by {\it modus tollens}).
\end{itemize}

\noindent Though Searle initially only asserts energy conservation as a matter of fact, the physical necessity ascription to conservation is inspired by Searle's claim that ``one of the most fundamental laws of physics'' is involved. It seems that it is the modal version alone that gives the physicalist hope for an {\it a priori} argument against dualism which avoids dealings with empirical science beyond whatever was involved in arriving at the conservation of energy in the 19th century, which had more to do with steam than with brains.

The formalized objection from energy conservation still has room for improvement. In particular, the global appeal to the whole universe in P2 seems unnecessary, at least if telekinesis is excluded or limited to some finite region and physics has no action at a distance (which indeed current physics does not in this sense).  A simple and plausible version of that assumption is to assume that dualist minds affect only brains. Then a conservation claim for brains should suffice for P2, if one can be suitably formulated, irrespective of what is happening on Mars or Alpha Centauri. In fact, the global version makes P2 more vulnerable, since there is considerable doubt among cosmologists whether the total amount of the universe's energy can even be defined (Peebles 1993, p.~139). If calculation is meaningful, the results sometimes disagree or give surprising results (e.g., Nester {\it et al.}~2008). Total energy could fail to be defined because there is just too much of it, giving an integral that diverges to infinity, or perhaps because energy is bound up with coordinate systems and hence cannot be globally integrated if the universe requires multiple coordinate systems.\footnote{This latter problem could be addressed, though, by using a background notion of covariant derivative, albeit nonuniquely (Sorkin 1988).} One therefore should not rely on global notions of energy conservation (E = constant) if more robust notions are available.

One might therefore suggest the modified version of the formalized argument:

\begin{itemize}

\vspace{-0.1cm}\item[P1']
If nonphysical minds influenced brains, then energy would not be conserved in brains.
\vspace{-0.2cm}\item[P2'] Energy is conserved in brains (with physical necessity). 
\vspace{-0.2cm}\item[C'] Therefore, dualism is false (by {\it modus tollens}).
\end{itemize}
The shift from phrases such as ``energy is constant'' to ``energy is conserved'' requires explanation. Constancy is straightforward: being the same over time.  For the whole physical universe, constancy might seem like an adequate concept (but note the above difficulties).  As regards the brain, however, energy obviously changes in it all the time; as a living organ, the brain is a swirling sea of metabolism and neural activity, and in permanent matter and energy exchange with its immediate physiological environment.  In other words, the brain is an open system. How can energy conservation be defined for an open system?

Here the fact (on which more below) that modern physics involves {\it local} conservation laws makes its first appearance. Local conservation laws describe conservation at every point in space, not (just?) for the world as a whole. Starting with local conservation laws and adding them up (integrating them) from different places, one gets a serviceable (albeit logically weaker) formulation suitable for finite regions such as brains. How can this local version of the principle of energy conservation be put in non-mathematical terms?  

A suitable formulation is Collins's (2008, p.~34) boundary principle of energy conservation (BPEC): 
\begin{quote} \small
According to the BPEC, the rate of change of energy (...) in a closed region of space is equal to the total rate of energy (...) flowing through the spatial boundary of the region.
\end{quote}
This principle is intended to be applicable to any and every region of space. Thus energy cannot disappear in Cambridge and immediately reappear in Liechtenstein, or disappear in Cambridge and reappear a bit later (due to the speed of light) in Liechtenstein, or just disappear in Cambridge, or just appear in Liechtenstein, etc., according to the BPEC and the local conservation laws that it paraphrases.  Thus, energy will be conserved in a brain in the relevant sense if the energy change in the brain can be accounted for by the amount of energy flowing through the brain's spatial boundaries.  If someone thinks that an immaterial mind can act on the whole body, then ``brain'' can be replaced with ``body''.  In summary, the BPEC claims that there must be physical causes which explain the energy change in an open system.

At this point a remark about quantum physics is in order.  With the advent of quantum mechanics and the Heisenberg uncertainty in particular, a certain ``blur'' arguably affects energy and momentum conservation at the microscopic level.  The Heisenberg uncertainty relation for quantum mechanics says, roughly, that the blur in the momentum and the blur in the position trade off such that their product is (at least) some constant value related to Planck's constant.  There is also an energy-time uncertainty relation. These uncertainties have been taken by Beck and Eccles (1992) to allow for mental interactions without energy expenditure.  

However, how these uncertainty relations fit with the exact conservation of energy and momentum in Feynman diagrams in quantum field theory is not entirely obvious.  Fully resolving such questions would require addressing the infamous measurement problem of quantum mechanics, which {\it prima facie} appears to call for partly replacing the (conservation-respecting) ordinary dynamics of quantum mechanics or quantum field theory with occasional collapse of the wave function.\footnote{We thank an anonymous referee for helping to clarify this point.} For our purposes it is not necessary to resolve such issues, because quantum mechanics, if anything, makes life easier for the interactionist dualist, not harder (see Sec.~6 for more on this). One therefore faces the energy conservation objection to interactionism in its strongest form by facing it in classical field theory and ignoring quantum physics. 

The view defended here is that the best response to the energy conservation objection -- the response that reflects an understanding of the relevant theoretical physics -- is what has been called the ``conditionality response'' (Pitts 2019a) that energy is conserved when and where minds do not act on bodies, but is not conserved when and where minds act on bodies.  
There are some works in the literature which have proposed the conditionality of conservation.  Descartes himself might have held such a view (as applied to the conservation of motion, Garber 1983a).  In the 18th century Knutzen and Crusius explicitly held it (Watkins 1995, 1998).  
Newton might well have held such a view (see McGuire 1968 and Dempsey 2006\,\footnote{We thank Steffen Ducheyne for the Dempsey reference. Dempsey distinguishes Newton's view from substance dualism, however; apparently it is mental and physical properties that interact.} for Newton's strong affirmations of mental causation); while he didn't take the conservation of energy/{\it vis viva} to hold generally anyway, momentum conservation would still be an issue. There seems to be little reason from physics and no reason from the mind to expect Newton's third law of motion (action-reaction) to hold in the mind-body problem.  Thus momentum conservation is a likely casualty of mental causation.  
Euler, who first formulated local conservation laws (Euler 1755/1757), was a staunch proponent of interactionism and vigorous opponent of Leibniz-Wolff pre-established harmony (Euler 1768-1772), which was often motivated by the conservation argument.  Euler, the leading physicist of the mid-18th century, can hardly have failed to understand the conservation argument, so presumably he simply was not bothered by non-conservation.  Indeed this conditionality view might have been the usual view following the defeat of pre-established harmony prior to Helmholtz's renewed claim of universal conservation even including the mind.  Clearly {\it if} the team involved Descartes, Newton and Euler as well as Knutzen and Crusius, it was in a strong position. 

While more recent interactionists have often aspired to preserve conservation, a few 20th and 21st century interactionist dualists (or temporary sympathizers) have offered the conditionality response as well (e.g., Ducasse 1951, pp.~240--242, Larmer 1986, Plantinga 2007, Rodrigues 2014, Lycan 2018), as have theists replying to an analogous objection against divine interaction (e.g., Larmer 2014).  However, reasonably detailed and physically well-informed accounts have been rare, two examples being Averill and Keating (1981) and Plantinga (2007). 
Averill and Keating (1981) hold that energy and momentum are conserved in a system so long as no external force acts on it, but since the dualistic mind exerts an external force on the brain, non-conservation is to be expected.  They also rightly talk of mental force but not mental energy (which faces problems pointed out below). However, Averill and Keating do not mention the converse Noether theorem or the locality of energy conservation and do not fully grasp the parallels between energy and momentum, which would have made their arguments even stronger.  Plantinga (2007) discusses the locality of conservation, infers from symmetries to conservation laws, and thus does not expect conservation where symmetries are violated, though without making a clear inference from the converse first Noether theorem. 

The conditionality response appears to be necessary (Pitts 2019a) also partly because of the lack of success of the Mohrhoff-Collins invocation (Mohrhoff 1997, Collins 2008, Collins 2011a) of general relativity on behalf of interactionist dualism. They claim that general relativity already excludes conservation laws for energy and momentum, so there is nothing left for interactionist dualism to ruin. But at least formally, general relativity certainly does have conservation laws (Bergmann 1958, Anderson 1967, Schmutzer 1972, Misner {\it et al.}~1973, p.~465), perhaps stronger ones than earlier theories have; it simply has been difficult to find a reasonable physical interpretation of these relations, which involve ``pseudotensors''.\footnote{A potential distraction is the zero {\it covariant} divergence of the {\it material} stress-energy-momentum tensor $\nabla_a T^{ab} = 0$, ignoring gravitational energy, perhaps due to its highly controversial status. This relation is true, important, and entailed by Einstein's field equations. However, it is not all that distinctive of Einstein's theory, because the material field equations of various other gravitational theories also imply this relation (Freund {\it et al.}~1969,
Wald 1984, p.~456).  The crucial point, however, is that typically $\nabla_a T^{ab} = 0$ simply does not imply the conservation of anything in the usual sense of total = constant over time because the covariant divergence cannot be integrated (Weyl 1922, pp.~236, 269--271, Misner {\it et al.}~1973, p.~465, Landau and Lifshitz 1975, p.~280, Lord 1976, p.~139, Stephani 1990, p.~141); neither does it imply the ``continuity equation'' describing local conservation laws, which will be discussed in detail below. Rather, $\nabla_a T^{ab} = 0$ is a balance equation saying at what rate material energy and momentum are produced/destroyed due to
gravitational influence. \\ Given that formally there are other conservation relations (involving a ``pseudotensor'' $t^{ab}$ of gravitational stress-energy-momentum along with the material stress-energy-momentum $T^{ab}$) that {\it do} involve the continuity equation and {\it do} imply total = constant over time, the pseudotensor relations are more relevant to the question at hand. Pseudotensor laws have been difficult to interpret physically and hence tend to get little attention, but they are still true and are the closest analogs of the conservation laws in earlier theories.  The formal pseudotensor conservation laws follow from Noether's theorem due to symmetries (uniformities of nature) and {\it vice versa} by the converse Noether theorem, whereas the relation of zero covariant divergence of the material stress-energy follows instead from matter's coupling only to a space-time metric and is not essentially connected to uniformities of nature.  A reasonable theory of gravity could fail to include $\nabla_a T^{ab} = 0$ if matter doesn't couple solely to a metric tensor (Blanchet 1992, Pitts 2016), so an attempted conservation objection to interactionist dualism using this relation would be both novel and unconvincing.  If one attempts to take Noether's (first) theorem seriously, one can view some of the peculiarities of pseudotensors as due to their describing infinitely many conserved energies and momenta, not just one and three (respectively), so it is unmotivated to expect distinct conserved quantities to be equivalent (Pitts 2010).
} 
Now one can avoid contestable interpretations and do new calculations showing that general relativity makes it harder, not easier, for souls to act on bodies (Pitts 2019b).  One can, if one interprets gravitational energy mathematics literally or realistically (Pitts 2010), view this new general relativistic objection as a strengthened energy conservation objection, but that is optional. 

We can now discuss several unsuccessful efforts to reconcile interactionist dualism with physical conservation laws.  All of them aim at refuting P1':  If nonphysical minds influenced brains, then energy would not be conserved in brains.

\bigskip
\bigskip

\noindent {\large\bf 2. Unsuccessful Dualist Responses I:} 

\vspace{0.1cm}
 \hspace*{0.1 cm}{\large\bf  Energy Redistribution vs.~Symmetry}

\medskip
This type of response claims that the mind does not add or subtract energy but {\it redistributes} the existing amount of energy. Thus, the hope is, the mind can interact with the brain while respecting energy conservation by deducting or otherwise compensating energy from elsewhere. Instances of this idea can be found, e.g., in the works of Broad (1937, p.~108), Dilley (2004, p.~142), Meixner (2008, p.~18) and Gibb (2010).  This proposal, though somewhat common, fails for more than one reason, and so will be addressed in this section and the next.

One problem is that it fails to recognize that conservation laws are consequences of symmetries by Noether's first theorem (Noether 1918, Goldstein 1980, Chap.~12-7). A continuous symmetry is described by a parameter that can be as small as you like, such as a translation by
1~m, or 1~mm, or $10^{-6}$~m, or ..., rather than by an essentially large transformation such as a reflection of a right hand into a left hand.
A symmetry such as time (or space) translation means that the laws do not treat any time (or place) as special, a kind of uniformity of nature. When there is a single independent variable (time), Noether's theorem leads to the constancy (zero time derivative, that is, zero rate of change over time) of the corresponding quantity. When there are multiple independent variables (time and space), Noether's theorem leads to the continuity equation, which relates the rate of change of 
the density of a quantity with the amount of the quantity spewing out (the divergence of the current density). 
A simple analogy would involve the number of persons in a room (assuming no conceptions/births or deaths): the number of persons changes only insofar as people enter or leave through the doors; people never disappear, pop into existence, or teleport into or out of the room. The mathematical form of the continuity equation will appear below. As noted above, in many circumstances one can integrate (add up) the continuity equation over ``all space'' and derive constancy of the total amount of the quantity in question (energy, momentum, angular momentum, or the like.) But the global form is not fundamental and might not exist mathematically (see Sec.~1). Integrating over some part of the universe also gives useful results.

Energy conservation follows from time translation invariance; momentum conservation follows from spatial translation invariance (Goldstein 1980).  In other words, if the physical laws are the same over time, then the amount of energy does not change over time; if the physical laws are the same across space, then the amount of momentum does not change over time. Hence nonphysical mental influence, being restricted in time and place to the times and places where minds act on brains, will violate the assumptions on which the conservation laws are based. Thus it might well be the case that energy and momentum are not conserved, if nonphysical minds act on brains. Note further that if such symmetries hold in some regions and fail in others, then the corresponding conservation laws will hold in the symmetric regions and fail in the non-symmetric regions. 
Symmetries need not apply to the entire universe to imply conservation laws, as will be evident from the calculations below. If souls exist only on the surface of the Earth while human beings exist, then conservation of energy and momentum apply everywhere away from the surface of the Earth while human beings exist, etc.  While energy conservation is often called the first law of thermodynamics, energy non-conservation due to souls does not need to worry engineers who design thermodynamic systems:  assuming that there are no souls in refrigerators, then energy is conserved there and the relevant (spatially localized) portion of the first law of thermodynamics still holds. Thus Bunge's catastrophe, in which ``physics, chemistry, biology, and economics would collapse'' due to energy non-conservation (Bunge 1980, p.~17), is a fantasy, absent some reason to think that the extent of non-conservation would be large. 

One can say more. Due to the converse of Noether's first theorem (Noether 1918, Brown and Holland 2004, Kosmann-Schwarzbach 2011), conservation laws imply symmetries, or contrapositively, non-symmetries (singling out some times and places over others) imply non-conservation. The Noether theorem with its converse entails a biconditionality between a continuous symmetry and a conserved quantity. 

\bigskip
\centerline{\bf Continuous Symmetry $\longleftrightarrow$ Conserved Quantity}

\newpage
\noindent Thus the action of a non-physical mind on the brain, assuming that the mind does not act in exactly that same way at every moment in time and every point in space throughout the entire universe (which seems a safe assumption), does indeed entail non-conservation of energy (by singling out some times as those when the mind acts on the brain) and non-conservation of momentum (by singling out some places as those where the mind acts, presumably within the brain only).\footnote{Here we are attempting to avoid ``violation'' language. Unlike philosophers, physicists routinely use the words ``violate'' and ``violation'' with little or no connotation of naughtiness or absurdity. Some physicists study CP-violation; some study Lorentz-violating theories of gravity. Talk of violation suggests some measure of surprise, but does not imply that the subject matter is non-existent, criminal, naughty, or highly implausible on balance. While it is useful to try to adapt one's usage to one's audience, philosophers also should learn not to think that they hear what isn't actually said by physicists due to differences in vocabulary.}
Interactionists need to accept this fact, not devise allegedly physics-respecting dodges that contravene basic theorems in real physics.  This non-conservation inference is not an objection to interactionism, however (or not much of one), because the hypothesis of interactionism defeats the presumption that one should be able to extrapolate inductively from conservation where minds do not act to conservation everywhere. It is a direct consequence of Lagrangian local field theory that if non-physical minds affect the physical world (as described by classical fields) in some times and places but not others, then minds produce and/or destroy energy and momentum at some times and places. If one wonders how an immaterial mind is able to produce or destroy energy, this question is not an appropriate candidate for an intuition because energy is a technical notion in physics. Either Lagrangian field theory gives the answer, or one has shifted from a physical objection involving energy to a Princess Elisabeth-style metaphysical objection that nonphysical minds ought not to be able to affect bodies (see Sec.~1).  

Typically one can read off momentum and/or energy conservation from the Lagrangian, the function that (using the principle of least action) fixes the equations of motion; in simple cases the Lagrangian is the kinetic energy {\it minus} the potential energy. If no spatial variable (usually $x$, $y$, $z$) figures in it (i.e., if the Lagrangian does not explicitly depend on place), momentum is conserved. If the Lagrangian depends on $x$ but not $y$ or $z$, then momentum in the $x$-direction is not conserved, but momentum is conserved in the $y$- and $z$-directions. Most of that (apart from the distinction of components in various directions) goes for energy as well; one just needs to replace ``spatial variable(s)'' by ``time variable'' (though the resulting constancy is still with respect to time).  On the other hand, if the Lagrangian explicitly depends on place, then momentum is not conserved.  Note that if one is dealing with particles rather than (as is standard in fundamental physics) fields, then {\it differences} of particle coordinates, such as $x_2 - x_1$, are permitted in the Lagrangian while still conserving momentum, because translating the whole universe will leave such differences unchanged:
$(x_2-c) - (x_1 - c) = x_2 - x_1$.

The physical objection against the idea of energy and/or momentum redistribution might suffice to refute that approach. However, there is a further difficulty for someone seeking to preserve energy conservation for interactive dualism in the way outlined:  it is exceedingly difficult to think of a plausible compensation mechanism in the first place. It isn't at all clear either how such a mechanism could work metaphysically or how it could be formulated so as to have an interface with physics. Such a mechanism will maintain or aggravate the violation of time- and/or space-translation invariance that, if not violated, would have ensured that conservation of energy and/or momentum. In terms of physics-friendliness, nothing is gained and much is lost, somewhat like treating measles by injecting influenza, or robbing Peter to repay Paul after robbing Paul.

In summary, redistribution approaches fail to achieve their goal, namely upholding conservation, because they fail to recognize that the nonphysical mind's action on the brain violates time- and space-translation invariance and hence, by the converse first Noether theorem, implies that energy and momentum conservation are violated in the sense relevant to modern physics.

\bigskip
\bigskip

\noindent {\large\bf 3. Unsuccessful Dualist Responses II:} 

\vspace{0.1cm}
 \hspace*{0.1 cm}{\large\bf  Energy Redistribution vs.~Locality}

\medskip

The proposal of energy redistribution to uphold conservation laws also displays a radical failure to understand the target, that is, what the conservation laws in modern physics actually are. The previous section to some extent played along with the assumption, typical in the philosophy of mind, that physics is about particles in discrete locations. But in modern physics, everything including such particle-like entities as protons and electrons, is an excitation of one or more continuous fields with local interactions. A field has one or more values at each time {\it and place}.\footnote{These values are the same in all coordinate systems for ``scalar'' fields, but are relative to coordinate systems in a rule-governed way for more complicated fields. There can also be other forms of conventionality in ``gauge theories''.} 

In such physics, the conservation laws are local. Imagine dividing the world into as many little mathematical boxes as you like with a space-time coordinate system. In each little box, the amount of energy remains constant except insofar as energy flows in or out through the box's (imaginary) walls. Thus conservation of energy is not one equation, but one equation {\it for each point in space} (Lange 2002, Chap.~5, Pitts 2019a), a sort of continuous conjunction. 
This fact brings us back to the BPEC, a formulation of local conservation laws: given any region of space, the change of energy within that region equals the flow of energy across the boundaries of that region of space.

Ducasse, having previously proposed a conditionality view that we find more reconcilable with physics (Ducasse 1951, Chap.~12), later proposed that (Ducasse 1960, p.~89) 
\begin{quote} \small
it might be the case that whenever a given amount of energy vanishes from, or emerges in, the physical world at one place, then an equal amount of energy respectively emerges in, or vanishes from, that world at another place.
\end{quote} 
The account is tantamount to a ``teleportation'' of energy from one place to another, which might happen simultaneously or with a time lag (due to the speed of light).  But both versions face the problem that according to modern physics, matter is continuous and acts locally through the propagation of waves, which means that conservation laws apply {\it locally} and that (at least assuming relativistic physics) a transport of energy can happen no faster than at the speed of light. The former version additionally runs into the difficulty that simultaneity is relative according to relativistic physics. 
To be sure, relativity might someday fall in favor of absolute simultaneity (cf.~Ho\v rava 2009), but even such a dramatic outcome would not change the locality of conservation as long as the successor theories are local field theories. Thus, redistribution fails to even recognize what the correct target concept of conservation is, to say nothing of hitting it.  The BPEC excludes any form of redistribution as a way to uphold conservation in the sense that matters in modern physics. Given the correct, that is local, form of conservation laws, compensation at a distance will exclude space translation invariance and hence will imply non-conservation of momentum both where the nonphysical mind acts and wherever any compensation occurs. Compensating for one violation with an additional one is hardly progress.

\bigskip\bigskip

\noindent {\large\bf 4. Unsuccessful Dualist Responses III:} 

\vspace{0.1cm}
 \hspace*{0.1 cm}{\large\bf Lowe's Alteration of Constants}

\medskip

	E.~J.~Lowe at some stage envisaged that the mind could act on the body by changing physical constants, which he took to be an energy-conserving proposal. A previous work noted that such a proposal would still imply non-conservation (Pitts 2019a), but left a more detailed critique of this proposal for the future. This paper takes up that issue. Lowe (1992, p.~270) wrote (emphasis in the original):
\begin{quote} \small
According to this second line of thought, the mind exerts causal influence on the body not through the exercise of psychic (non-physical) {\it forces} of any sort, but through influencing the values of the so-called ``constants'' which feature in various {\it physical} force laws -- for instance, by influencing (presumably only locally and to a vanishingly small degree) the value of the universal constant of gravitation $G$ or the value of the charge on the electron. Thus it would turn out that these so-called ``constants'' are strictly speaking {\it variables}. 
\end{quote}

Lowe envisaged that if the mind were able to change certain physical constants, such as, e.g., the gravitational constant $G$ or the value of the electron charge $e$, it could influence the brain without adding to or subtracting energy from it. He admits that ``postulating variability in such constants is in a sense at odds with the classical principle of the conservation of energy'' (Lowe 1992, p.~270), but still prefers this to accepting the violation of energy conservation, because ``it is not now being suggested that the mind has a power of creating energy {\it ex nihilo} or conversely annihilating it'' (Lowe 1992, p.~270).  This latter statement makes it clear that he believes that a change in physical constants does not result in energy being added or subtracted.

	Unfortunately, Lowe's theory does not yield his desired result of energy conservation. To see this, consider the following model calculation, which is perhaps the simplest possible example representative of modern physics. This calculation is intended not merely to refute Lowe's proposal, but also to make as widely available as possible the physics of conservation laws, a topic often discussed but not often understood correctly in the philosophy of mind. 

	Suppose a physical system on which the mind acts is a massive scalar field $\phi(t,x)$  (in one spatial dimension for ease of calculation). A scalar field is one number at each point in space and time; it is called a {\it scalar} field because the number is the same in every coordinate system.\footnote{Most realistic fields are more complicated; for example, the electromagnetic field comes from a 4-component potential, leaving one with 4 fields (the values of which depend on the coordinate system), as well as four independent variables $t$, $x$, $y$, and $z$ in real 3-dimensional space. The toy calculation presented includes everything relevant to understanding the more realistic calculation with more fields and more spatial dimensions, while omitting a great deal of irrelevant complexity.} In the example calculation, this field stands in for all the usual physical fields (electromagnetism, gravity, the weak and strong nuclear forces, the electron field, neutrino fields, etc., nearly all of which are more complicated than scalar fields), while the mass parameter $m$ stands in for whatever physical constants Lowe envisages as varying due to mental influence. 

The Euler-Lagrange field equation (also known as equation of motion) follows from the principle of least action, which says roughly that the action, a time integral of the Lagrangian, is as small as possible, given the beginning and ending configurations. The Euler-Lagrange equations are second-order partial differential equations whose solutions are the functions for which a given functional, the action (the time integral of the Lagrangian), is stationary, i.e., the system's action is ``least'' or perhaps ``most'' or at any rate unchanged by a small change in the dynamical variables. The Lagrangian is basically the kinetic energy {\it minus} the potential energy. For a local field theory (Goldstein 1980, Chap.~12), the Lagrangian is given by a Lagrangian density  by adding up (integrating) the Lagrangian density over all points of space (here represented by $x$ because there is assumed to be only one spatial dimension):

\begin{equation} L = \int_{-\infty}^{+\infty} dx \, \cal L 
\end{equation}
The Lagrangian density for this field is
\begin{equation}
{\cal L} = {1\over 2} \dot \phi^2- {1\over 2} \phi'^2 - {1\over 2} \, k(t,x) \, \phi^2
\end{equation} 
where $\dot \phi$ is the partial time derivative $\partial \phi / \partial t$ (describing how $\phi$ changes over time at constant location $x$), $\phi'$ is the partial spatial derivative $\partial \phi / \partial x$ (describing how $\phi$ changes from place to place at constant time $t$), and $k$ is the square of the ``mass'' of the field.\footnote{Actually $k$ must be the square of an inverse length; one converts such a thing to a mass using the speed of light and Planck's constant. The details are not important here. Here such constants are set to the value of 1; one could set the speed of light $c$ to 1 by measuring time in years and distance in lightyears, for example. The term ``mass'' is used because if one were to quantize this field, it would lead to quanta with mass $\sqrt k$ at least if $k$ were constant.}

Here $k$ is allowed to vary in place and time to represent the influence of Lowe's constants-altering mind. One uses this Lagrangian density to infer the equation of motion or field equation for this field, the Euler-Lagrange equation for that field. The Euler-Lagrange equation in some cases takes the form $-ma + F = 0$ and thus can be a mere rearrangement of Newton's second law, but the Lagrangian formalism is more general in some respects and is the standard starting point for fundamental physics. One can instantly see from this Lagrangian that energy is not conserved in this case, because the Lagrangian explicitly depends on time due to $k(t,x)$. The details of the calculation are useful, however, in making as accessible and simple as possible the mathematics-physics that is usually missing in discussions of conservation laws in the philosophy of mind.

The principle of least action implies the Euler-Lagrange equation (ELE) for the above system, 
$$ {\partial {\cal L} \over \partial \phi} - {d\over dt}  {\partial {\cal L} \over \partial \dot \phi} -
{d\over dx}  {\partial {\cal L} \over \partial \phi'} = 0 \,. $$
With the Lagrangian density from above substituted, the ELE ends up reading
\begin{equation}
-k(t,x) \phi - \ddot{\phi} + \phi'' = 0 \,.
\end{equation} 
In order for energy conservation to hold in such a system, the following continuity equation must hold (Goldstein 1980, Chap.~12):
\begin{equation}
{d\over dt} \theta_0^0 + {d \over dx} \theta_0^1 = 0
\end{equation} 

This form of equation\footnote{Unfortunately, there is no clear notation for these expressions, because neither the partial derivative $\partial$ nor the ordinary derivative $d$ notation is quite clear. In taking, for example, the time derivative of the energy density, one wants all dependence on time, whether implicitly through the fields and their derivatives, or explicitly through the appearance of $t$ itself (though that would be peculiar in this context), but none of the dependence on space.} 
also describes the conservation of charge (Griffiths 1989, p.~4). Returning to the case at hand, $ \theta_0^0$ and  $\theta_0^1$  are components, the energy density and energy flux density, respectively, of the canonical energy-momentum tensor,
\begin{equation}
\theta_\nu^\mu = {\partial {\cal L} \over \partial \left({\partial \phi \over \partial x^\mu}\right)} \,  {\partial \phi \over \partial x^\nu} - \delta_\nu^\mu {\cal L} \, ,
\end{equation} 
with $\mu$ and $\nu$ both ranging from 0 to 1 and $\delta$ (the Kronecker $\delta$) being 0 if $\mu \neq \nu$ and 1 if  $\mu = \nu$. (It should be added that $x^0$ is identical to $t$ (time) and $x^1$ to $x$ (the only spatial coordinate).) Thus, we obtain the energy density
\begin{equation}
\theta_0^0 \, = \, {\partial {\cal L} \over \partial \left({\partial \phi \over \partial t}\right)}\,  {\partial \phi \over \partial t} - {\cal L} \, = \, {1\over 2} \dot \phi^2 + {1\over 2} \phi'^2 + {1\over 2} \, k(t,x) \, \phi^2
\end{equation}
and the energy flux density:
\begin{equation}
\theta_0^1 \, = \, {\partial {\cal L} \over \partial \left({\partial \phi \over \partial x}\right)}\,  {\partial \phi \over \partial t}
- 0 = - \phi' \dot \phi 
\end{equation}
If one plugs the above expressions (6) and (7) into (4), one gets:
\begin{eqnarray} 
{d\over dt} \left({1\over 2} \dot \phi^2 + {1\over 2} \phi'^2 +  {1\over 2} \, k(t,x) \, \phi^2 \right) + {d\over dx} (-\phi'\dot\phi) &=&   \nonumber \\
\ddot{\phi}\dot\phi + {\boldsymbol{\dot\phi' \phi'}}  + {1\over 2} \left(\phi^2 {\partial \over \partial t} k(t,x) + 2k(t,x) \dot\phi \phi \right)    - \phi'' \dot\phi - {\boldsymbol{\dot\phi' \phi'}}        &=& 0 \hspace{0.3cm} \nonumber
\end{eqnarray}
The terms in bold cancel out, and application of the Euler-Lagrange equation (3) yields:
\begin{eqnarray} 
(-k(t,x) \phi - \ddot{\phi} + \phi'')(-\dot\phi) + {1\over 2} \phi^2 {\partial \over \partial t} k(t,x) &=& 0 \, , \nonumber \\
0 +  {1\over 2} \phi^2 {\partial \over \partial t} k(t,x) &\neq& 0 \, . \nonumber
\end{eqnarray}

In other words, the continuity equation (4) does not hold for energy just in case $k$ varies with time. Hence, precisely because of Lowe's assumption that the mind can alter physical constants, energy is not conserved in this system! The crux is the term which represents Lowe's alterable constant $k(t,x)$. Its time derivative does not become zero or cancel out during the calculations. Thus one sees in the breach how time translation symmetry implies the conservation of energy, because here the failure of that symmetry implies non-conservation. 

Might at least momentum be conserved, as Lowe  (1992, p.~268) seems to assume? Since the Lagrangian explicitly depends on place (due to $k(t,x)$), momentum cannot be conserved either. Mathematically, momentum conservation can be checked by the corresponding continuity equation involving the momentum density
$$
\theta_1^0 \, = \, {\partial {\cal L} \over \partial \left({\partial \phi \over \partial t}\right)}\,  {\partial \phi \over \partial x} - 0 = \dot \phi \phi' $$
and momentum flux density
$$\theta_1^1 \, = \, {\partial {\cal L} \over \partial \left({\partial \phi \over \partial x}\right)}\,  {\partial \phi \over \partial x} - {\cal L} \, = \, {1\over 2} \dot \phi^2- {1\over 2} \phi'^2 + {1\over 2} \, k(t,x) \, \phi^2 \, :
$$
\begin{equation}
{d\over dt} \theta_1^0 + {d\over dx} \theta_1^1 = 0
\end{equation}
Running the above calculations on (8) yields
\begin{equation}
(-k(t,x) \phi - \ddot{\phi} + \phi'')(-\phi') + {1\over 2} \phi^2 {\partial \over \partial x} k(t,x) = 0 \, ,
\end{equation}
which, after use of (3), results in:
\begin{equation}
0 + {1\over 2} \phi^2 {\partial \over \partial x} k(t,x) \neq 0 \
\end{equation}

Thus, momentum is not conserved either, precisely due to the spatial variation of the erstwhile constant. One also sees that whenever and wherever the ``constant'' is really constant, energy and momentum are conserved, even if conservation does not hold in some regions, such as in brains. To the trained eye, key qualitative features of these calculations were evident in advance of doing the mathematics. But working the example is useful for those not antecedently interested in classical field theory. This mathematics shows what training the eye needs to see the failure of Lowe's varying-constants proposal as well as why non-conservation in some regions doesn't spoil conservation elsewhere.

\bigskip
\bigskip

\noindent {\large\bf 5. Problematic Dualist Response:} 

\vspace{0.05cm}
 \hspace*{0.1 cm}{\large\bf Ascribing Energy to the Soul}

\medskip
Another proposed way of constructing energy-conserving dualistic interaction is to claim that the mind carries energy and that an increase of energy in the body would be compensated by a corresponding decrease in the mind and {\it vice versa}, so that no violation of energy conservation takes place. Hart (1994, p.~268) explicitly offers such an account:
\begin{quote} \small
Energy (or mass-energy) is conserved, and conservation is a quantitative principle. So we need intrinsically psychological quantities. (...) Once we have such psychological quantities, we may imagine that as light from objects seen reaches the region of convergence along the disembodied person's lines of sight, it passes straight through but loses some electromagnetic energy and, at a fixed rate of conversion, that person acquires or is sustained in visual experience of those objects seen. (...) So we have solved the interaction problem. To be sure, we have not imagined exactly how light energy converts into the psychic energy implicit in visual experience. But then neither do physicists tell us how mass turns into energy when an atom bomb goes off, and if their lacuna does not embarrass them, neither need ours embarrass us. 
\end{quote}

Searle used similar talk of mental energy to argue that dualism violates energy conservation (see Sec.~1). He identifies the problem as mental energy being ``not fixed by physics'' and hence not figuring in the ``energetic bookkeeping'' of physics. As noted in Sec.~1, Searle didn't explicitly say that physical and mental energy are interconvertible. But does convertibility of physical and mental energy, as suggested by Hart, really help much? 

	Does mental energy even make sense, given that (to avoid mere equivocation on the word ``energy'') mental energy would involve mathematical properties of the non-physical mind? Which rate of change of a given belief or desire offsets a speed of 2 meters per second for a 1 gram mass? In the early 18th century it was often said that there was no proportionality between mind and matter. Absent a treatment of the mind itself (not just interaction with the brain) in terms of Lagrangian field theory, we are unable to understand an ascription of energy to the mind in any sense relevant to the conservation objection. But a treatment of the mind in terms of Lagrangian field theory would make mental operations not merely deterministic (which one might or might not accept), but deterministic in a sense that involves causes that seem to have little or no connection to reasons, beliefs, desires, etc. Absent the extraordinary accomplishment that would be involved in addressing this objection, ascribing energy to the mind appears hopeless at least as a way to uphold conservation laws. 

Regarding Hart's parity claim for mass-energy conservation in a nuclear bomb, physics gives a detailed quantitative trade-off of two forms of energy, mass-energy of some nuclei made of protons and neutrons, and electromagnetic energy, all described in terms of various particles that
satisfy the relativistic equation $E = \sqrt{m^2c^4 + p^2c^2}$ (with $m$ varying with the particle type and presumably being 0 for photons).\footnote{This expression has the perhaps more familiar non-relativistic limit  $E= mc^2 + p^2/2m$ using the binomial series expansion; dropping the rest-mass energy gives, after expressing momentum in terms of velocity, $mv^2/2$. It is not actually required that photons have zero mass; the problem of the photon mass gives a physically interesting example of (perhaps) permanent underdetermination from approximate but arbitrarily close empirical equivalence (Pitts 2011).} There is no dearth of intelligibility regarding conservation in bombs. It is difficult to see what Hart could want that physics does not supply, unless one is disturbed that quantum mechanics seems not to be deterministic (in which case one can satisfy oneself with a deterministic interpretation) or one laments the essentially non-perturbative nature of quantum chromodynamics, which makes it difficult to cash out verbal claims about nuclei as composed of protons and neutrons, and protons and neutrons as composed of quarks. By contrast, Hart's proposal of mental energy is unintelligible apart from a treatment of the mind in terms of Lagrangian field theory, which seems quite unpromising. There is no parity between the two cases. 

Even if one were prepared to ascribe energy to the soul and if some exchange rates between mental and physical states were somehow stipulated, it is still by no means clear that the result would be that energy is conserved. Merely having exchange rates does not suffice, as anyone who goes to an airport and finds MoneyCorp buying dollars for a certain number of pounds sterling and selling dollars for a very different number of pounds will see. More seriously, both daily experience and (for those who embrace it) libertarian freedom make it highly implausible that the trajectory of thoughts will reliably harmonize with the trajectory of physical matter so as to ensure conservation. 

To our knowledge, the only philosopher who has seriously attempted to ascribe mathematical properties to the soul is Robin Collins. To be sure, his ``dual-aspect model of the soul'' (Collins 2011b) is not designed to answer the objection from energy conservation (for which he offers a different answer). His idea is that the soul has two kinds of properties, subjective (or mental) and non-subjective (or physical) properties, thus making the soul, on his definitions, a physical entity with additional subjective properties. He defines a physical entity as one whose ``states can be described by some mathematical function'' and the evolution of whose states and their interaction with other physical systems ``can be specified by a set of mathematical equations'' (Collins 2011b, p.~234). 

How do the mental and physical aspects of the soul work together? Collins proposes ``linking laws'' between mental states/qualia and physical states within the soul, e.g., a certain vibrational pattern (brought about by some specific brain activity) always produces a certain mental state/quale, and {\it vice versa}. Be that as it may, this proposal, if mental-to-physical causation is permitted, does not uphold the relevant conservation laws of physics (assuming that physics is described by the principle of least action or is even sufficiently well anticipated by that principle, as quantum field theory, our best standard theory, is). 

The symmetry-conservation law link (Noether's first theorem and its converse) implies that the conservation laws will hold if and {\it only if} the laws are the same everywhere and always, which can be true only if the soul either does nothing (or does nothing that wouldn't have happened anyway, which looks like nothing in the physics), or does the same thing everywhere and always throughout space-time. Collins admittedly has his own response to the conservation law objection by appeal to the supposed non-conservation of energy in general relativity (Collins 2008, Collins 2011a). Unfortunately, as noted above, this response does not work because general relativity in fact makes the situation for mind-to-body causation harder, not easier (Pitts 2019b). 

\bigskip\bigskip

\noindent {\large\bf 6. The Optimal Dualist Response?} 

\vspace{0.1cm}
 \hspace*{0.1 cm}{\large\bf Conditionality of Energy Conservation}

\medskip
We take it that the best reply a dualist can give (perhaps excepting some essentially quantum mechanical approach) is the conditionality (or biconditionality) response: energy is conserved in a system {\it on the condition} that no mind (or other non-physical entity, for that matter) acts on the system; conversely, {\it on the condition} that a mind acts on the system, energy is not conserved (Pitts 2019a). Whereas most of the other views above are supposed to respect physics but in fact are in mathematical conflict with modern physics, this view, which is often taken to conflict with physics (such as by Dennett and Bunge), is at least in accord with modern physics in the sense of being an option allowed by Noether's first theorem and its converse. However, it should be made clear, the conditionality response does not require any ascription of mathematical properties to the soul (though it does of course ascribe mathematical properties to the soul's influence on physical fields; otherwise one could not even discuss the conservation question in any serious fashion); at any rate, we reject such an approach (see Sec.~5).

There seems to be nothing left of the energy conservation objection, once one embraces the relevant mathematics of Lagrangian field theory and Noether's theorem and its converse, that isn't simply a form of metaphysical objection about a causal nexus or an open begging of the question (the problem with having souls act on bodies is that then souls act on bodies). If interactionism is under discussion, then there simply isn't a strong presumption in favor of exact energy conservation in the brain that has a claim on interactionists, except insofar as one has looked at the brain -- which makes it less of a presumption than an empirical result. 

Having confidence that one can extrapolate conservation laws into the brain is just having confidence in some form of physicalism, Leibnizian pre-established harmony, epiphenomenalism, or some other non-interactionist view. If interactionism weren't motivated by arguments, then refusal to extrapolate conservation into the brain would be unreasonable, like thinking that conservation holds everywhere except on Venus. But interactionism is in fact motivated by arguments, such as (e.g.) that causal closure is a self-defeating belief (Swinburne 2018), so this refusal to extrapolate to exact conservation in the brain is not frivolous. Looking at the brain, especially if one knows how carefully to look, should be decisive or at least highly constraining. 

Recall that Noether's first theorem has instantiations such as that energy is conserved on the condition that a system evinces translational symmetry over time (likewise momentum and space). The interactionist dualist can plausibly suggest that mental interactions simply do not conserve energy, because they break the temporal symmetry of the pertinent system, i.e.~the brain.\footnote{``Breaking'' a symmetry, much like ``violating'' a conservation law, is not necessarily bad.}
 Such a claim might seem disturbing if one takes conservation laws as a black box received on the authority of Science, as is customary in the philosophy of mind and metaphysics. But it will seem natural, even obvious, if one frames the question in terms of the real physics of fields (or its warmup exercise in terms of classical fields, rather) and contemplates Noether's first theorem. In other words, if the mind makes it the case that the brain's Lagrangian (or the Lagrangian of whatever system the mind acts on) explicitly depends on time, then energy is not conserved (see also the analysis of Lowe's proposal above). 

Scientists and engineers routinely study systems that have time- and space-dependence; simply by not studying the entire universe, one introduces time- and space-dependence into the system of interest because of the influence of everything outside the system. While such considerations help to show that the conditionality response is possible, the converse first Noether theorem makes this conditionality response obligatory (at least within the confines of Lagrangian field theory): conservation implies symmetry, so equivalently non-symmetry implies non-conservation. Thus if Susie decides to raise her arm and her soul acts on nerves in her brain to start the relevant causal chain, then the time- and space-dependence of her soul's influence (in her brain and not on the Moon, during her lifetime and not before or after) implies the non-conservation of energy and momentum in her brain at that time. So what?
 
In effect, this reply rejects P2' as insufficiently supported by physics and dialectically unavailable in this context. Ironically, it is exactly this premise which detractors of dualism have taken to be a cornerstone of physics and which many dualists have striven to uphold in order to respect science. But submitting to physics is a good idea only if one correctly describes what physics actually says. Otherwise one is doing the philosophy of A-level (secondary school) chemistry, not the philosophy of physics (to recall a warning by Ladyman {\it et al.}~2007, p.~24). 

The Noether-inspired (bi)conditionality response implies that anti-dualists can no longer employ armchair {\it arguments} from energy conservation against interactionism, because the distance between the premise and the conclusion has disappeared. One could still try to complain that energy isn't conserved given interactionism, but it is unclear why this is an objection. It seems to amount to saying this: the problem with having souls act on bodies is that then souls act on bodies. Restating the interactionist view isn't much of an objection to interactionism, at least to those who sympathize with interactionism in the first place. Physics often and successfully deals with systems in which the equations have time- and/or space-dependence, so why not have one more example? 

Note that recognizing (bi)conditionality does not necessarily preclude every objection from energy conservation against dualism; one could look at the brain and then claim that, as a matter of fact, energy does seem to be conserved and hence interactionism is false. However, it is doubtful that framing a neuroscientific objection to interactionist dualism in terms of conservation laws would ever be profitable, because access to (non)conservation is quite indirect, involving calculating a function of derivatives of another function, as appeared above. Rather, one could simply observe that empirical neuroscience leaves insufficient room for a soul to act on the brain in the right sort of way. 

What about quantum-mechanical approaches of mind-brain interaction, towards which there is a commendable trend in recent dualistic literature (e.g., Corradini and Meixner 2014, Halvorson 2011)? Presumably a quantum model is the best approach for anyone aiming to give a defense or positive account of interactionist dualism, something that we are by no means attempting here. Also, quantum approaches might become interesting for dualists should energy turn out to be {\it in fact} conserved in brains, for then quantum physics would seem to be the only way to bridge mind and brain in a potentially energy-conserving manner. However, {\it one} of us takes the view that a closer look at the neuroscience of volitional action suggests that there probably is a conspicuous gap in the chain of bio-physical causes, a gap that might well be filled by an immaterial mind (Cucu 2019). If experimental bounds from neuroscience sufficiently constrain any putative nonphysical mental influence describable classically, {\it then} a quantum approach might be required if interactionist dualists are to address the resulting {\it a posteriori} objection.

In light of all this, we take the conditionality response to be helpfully robust. To be sure, the conditionality response is compatible with quantum mechanics; it just does not need quantum mechanics. First, conservation has its best chance of posing an objection to interactionism in precisely the context faced here, namely, classical field theory; since, as we claim to have shown, no {\it a priori} objection can be upheld in light of Noether's first theorem, the conditionality response is stable in so far as it is fully in accord with Noether's theorem. Second, the conditionality response does not depend on specific calculations that might be questioned, such as the account of Beck and Eccles (1992; see Wilson 1999 for a critique). Third, it is not subject to debates about the right interpretation of the underlying physics. Some claim that the measurement problem in quantum mechanics necessarily involves the mind (e.g., Wigner 1967, Schwartz {\it et al.}~2005, Halvorson 2011); many others ignore such an approach completely and focus on purely physical interpretations of quantum physics. At any rate, coming up with a good quantum model is quite another matter from coming up with some quantum model. No such interpretive issues arise in classical mechanics and field theory, so the treatment above will remain available. 

	It is also worth recalling that some physicists, aiming to address the measurement problem in quantum mechanics, already accept energy non-conservation even apart from the philosophy of mind (e.g., Bassi {\it et al.} 2013),\footnote{We thank an anonymous referee for this reference.} as Butterfield noted earlier.  Such examples show that experts are not nearly as threatened by the prospect of energy non-conservation as many philosophers of mind. Conservation laws are useful accounting devices and sanity checks, but it is not considered absolutely obligatory that they hold exactly and at all times and places.   
\bigskip
\bigskip

\noindent {\large\bf 7. Conclusion}

\medskip
The objection from energy conservation against interactionist dualism relies on the premises that (1) interactionist dualism entails that energy is not conserved in brains and (2) energy is (physically) necessarily conserved in brains. In contrast to 18th century, at least since the later 19th century, interactionist dualists have sought to avoid premise (1) and uphold premise (2) to make dualism conservation-friendly. These proposals fail in multiple ways. Special attention has been given to a proposal by Lowe, which in fact entails the non-conservation of both energy and momentum, contrary to his intentions. The best account in case energy is {\it in fact} not conserved in brains is the conditionality response, which derives directly from the first Noether theorem. Should energy turn out to be conserved exactly, essentially quantum-mechanical approaches might be the only chance for interactionist dualism.

\bigskip
\bigskip

\noindent {\large\bf Acknowledgments}

\medskip
This work was funded by grants from the John Templeton Foundation: grant \#59226 (A.C.C.) and grant \#60745 (J.B.P.). Two referees made helpful comments regarding content and exposition. 

\bigskip
\bigskip

\noindent {\large\bf References}

\medskip

{\small

\noindent Anderson J.L.~(1967): {\it Principles of Relativity Physics}, Academic, New York. 

\smallskip
\noindent Averill E.~and Keating B.F.~(1981): Does interactionism violate a law of classical physics? {\it Mind} {\bf 90}, 102--107.

\smallskip
\noindent Bassi A., Lochan K., Satin S., Singh T.P., and Ulbricht H.~(2013): Models of wave-function collapse, underlying theories, and experimental tests. {\it Reviews of Modern Physics} {\bf 85}, 471--527. 

\smallskip
\noindent Beck F.~and Eccles J.C.~(1992): Quantum aspects of brain activity and the role of consciousness. {\it Proceedings of the National Academy of Science of the USA} {\bf 89}, 11357--11361. 

\smallskip
\noindent Bergmann P.G.~(1958): Conservation laws in General Relativity as the generators of coordinate transformations. {\it Physical Review} {\bf 112}, 287--289. 

\smallskip
\noindent Blanchet L.~(1992): A class of nonmetric couplings to gravity. {\it Physical Review Letters} {\bf 69},  559--561.

\smallskip
\noindent Broad C.D.~(1937): {\it The Mind and its Place in Nature}, Kegan Paul, Trench, Trubner and Co., London.

\smallskip
\noindent Brown H.~and Holland P.~(2004): Dynamical versus variational symmetries: Understanding Noether's first theorem. {\it Molecular Physics} {\bf 102}, 1133--1139.

\smallskip
\noindent Bunge M.~(1980): {\it The Mind-Body Problem: A Psychobiological Approach}, Pergamon, Oxford.

\smallskip
\noindent Butterfield J.~(1997): Quantum curiosities of psychophysics. In {\it Consciousness and Human Identity}, ed.~by J.~Cornwell, Oxford University Press, Oxford, pp.~122--159. 

\smallskip
\noindent Collins R.~(2008): Modern physics and the energy-conservation objection to mind-body dualism. {\it American Philosophical Quarterly} {\bf 45}, 31--42. 

\smallskip
\noindent Collins R.~(2011a): The energy of the soul. In {\it The Soul Hypothesis: Investigations into the Existence of the Soul}, ed.~by M.C.~Baker and S.~Goetz, 
Continuum, New York, pp.~123--133.

\smallskip
\noindent Collins R.~(2011b): A scientific case for the soul. In {\it The Soul Hypothesis: Investigations into the Existence of the Soul}, ed.~by M.C.~Baker and S.~Goetz, 
Continuum, New York, pp.~222--246.

\smallskip
\noindent Cornman J.W.~(1978): A nonreductive identity thesis about mind and body. In {\it Reason and Responsibility: Readings in Some Basic Problems of Philosophy}, ed.~by J.~Feinberg, Dickenson Publishing Company, Encino, pp.~272--283.

\smallskip
\noindent Corradini A.~and Meixner U., eds.~(2014): {\it Quantum Physics Meets the Philosophy of Mind}, Walter de Gruyter, Berlin.

\smallskip
\noindent Crane T.~(2001): {\it Elements of Mind: An Introduction to the Philosophy of Mind}, Oxford University Press, Oxford.

\smallskip
\noindent Cucu A.C.~(2019): No perfect pass: How the energy conservation objection against dualism turns out to be physicalism's own goal. Preprint.

\smallskip
\noindent Dempsey L.~(2006): Written in the flesh: Isaac Newton on the mind-body relation. {\it Studies in History and Philosophy of Science} {\bf 37}, 420--441.

\smallskip
\noindent Dennett D.~(1991): {\it Consciousness Explained}, Little, Brown and Co., Boston.

\smallskip
\noindent Dilley F.B.~(2004): Taking consciousness seriously: A defense of Cartesian dualism. {\it International Journal for Philosophy of Religion} {\bf 55}, 135--153.

\smallskip
\noindent Ducasse C.J.~(1951): {\it Nature, Mind, and Death}, Open Court, La Salle.

\smallskip
\noindent Ducasse C.~(1960): In defense of dualism. In {\it Dimensions of Mind: A Symposium}, ed.~by S.~Hook,  New York University Press, New York, pp.~85--90.

\smallskip
\noindent Euler L.~(1755/1757): Principes g\'en\'eraux du mouvement des fluids. {\it  M\'emoires de l'acad\'emie des sciences de Berlin} {\bf 11}, 274--315. Reprinted in {\it Opera Omnia Series 2} {\bf 12},  54--91. Translated by T.E.~Burton and U.~Frisch as ``General principles of the motion of fluids'', {\it Physica D} {\bf 237}, 1825--1839 (2008). 

\smallskip
\noindent Euler L.~(1768--1772): Lettres \`a une princesse d'Allemagne sur divers sujets de physique \& de philosophie. {\it Imprimerie de l'Academie imp\'eriale des sciences}, Saint Petersbourg. Translated in part by H.~Hunter as ``Letters of Euler on Different Subjects in Natural Philosophy: Addressed to a German Princess; with Notes, and a Life of Euler'', by David Brewster. Harper, New York (1840).

\smallskip
\noindent Fodor J.~(1998): The mind-body problem. In  {\it Philosophy Then and Now}, ed.~by N.S.~Arnold, T.M.~Benditt, and G.~Graham, Blackwell, Malden, pp.~63--77. Reprinted from {\it Scientific American}, January 1981, 114--123.

\smallskip
\noindent Freund P.G.O., Maheshwari A., and Schonberg E.~(1969): Finite-range gravitation. {\it Astrophysical Journal} {\bf 157}, 857--867.

\smallskip
\noindent Garber D.~(1983a): Mind, body, and the laws of nature in Descartes and Leibniz. {\it Midwest Studies in Philosophy} {\bf 8}, 105--134.

\smallskip
\noindent Garber D.~(1983b); Understanding interaction: What Descartes should have told Elisabeth. {\it Southern Journal of Philosophy} {\bf 21}, S15--S32. 

\smallskip
\noindent Gibb S.~(2010): Closure principles and the laws of conservation of energy and momentum. {\it Dialectica} {\bf 64}, 363--384.

\smallskip
\noindent Goldstein H.~(1980): {\it Classical Mechanics}, Addison-Wesley, Reading.

\smallskip
\noindent Griffiths D.G.~(1989): {\it Introduction to Electrodynamics}, Prentice Hall, Englewood Cliffs.

\smallskip
\noindent Halvorson H.~(2011): The measure of all things: Quantum mechanics and the soul. In  {\it The Soul Hypothesis: Investigations into the Existence of the Soul}, ed.~by M.C.~Baker and S.~Goetz, 
Continuum, New York, pp.~138--163.

\smallskip
\noindent Hart W.D.~(1988): {\it The Engines of the Soul}, Cambridge University Press, Cambridge.

\smallskip
\noindent Ho\v rava P.~(2009): Quantum gravity at a Lifshitz point. {\it Physical Review D} {\bf 79}, 084008.

\smallskip
\noindent Kosmann-Schwarzbach Y.~(2011): {\it The Noether Theorems: Invariance and Conservation Laws in the Twentieth Century}, Springer, New York. Translated by B.E.~Schwarzbach.

\smallskip
\noindent Ladyman J., Ross D., Spurrett D., and Collier J.~(2007): {\it Every Thing Must Go: Metaphysics Naturalized}, Oxford University Press, Oxford. 

\smallskip
\noindent Landau L.D.~and Lifshitz E.M.~(1975): {\it The Classical Theory of Fields}, Pergamon, Oxford.  Translated by M.~Hamermesh.

\smallskip
\noindent Lange M.~(2002): {\it An Introduction to the Philosophy of Physics: Locality, Fields, Energy, and Mass}, Wiley, Malden.

\smallskip
\noindent Larmer R.~(1986): Mind-body interaction and the conservation of energy. {\it International Philosophical Quarterly} {\bf 26}, 277--285.

\smallskip
\noindent Larmer R.A.~(2014): Divine intervention and the conservation of energy: A reply to Evan Fales. {\it International Journal for Philosophy of Religion} {\bf 75}, 27--38.

\smallskip
\noindent Leibniz G.W.~(1985): {\it Theodicy: Essays on the Goodness of God and the Freedom of Man and the Origin of Evil}, Open Court, La Salle. Translated by E.M.~Huggard.

\smallskip
\noindent Lord E.A.~(1976): {\it Tensors, Relativity and Cosmology}, Tata McGraw-Hill Publishing, New Delhi. 

\smallskip
\noindent Lowe E.J.~(1992): The problem of psychophysical causation. {\it Australasian Journal of Philosophy} {\bf 70}, 263--276.

\smallskip
\noindent Lycan W.G.~(2018): Redressing substance dualism. In {\it The Blackwell Companion to Substance Dualism}, ed.~by J.J.~Loose, A.J.L.~Menuge, and J.P.~Moreland, Wiley Blackwell, Oxford, pp.~22--39.

\smallskip
\noindent McGinn C.~(1999): {\it The Mysterious Flame: Conscious Minds in a Material World}, Basic Books, New York.

\smallskip
\noindent McGuire J.E.~(1968): Force, active principles, and Newton's invisible realm. {\it Ambix} {\bf 15},  154--208. 

\smallskip
\noindent Meixner U.~(2008): New perspectives for a dualistic conception of mental causation,. {\it Journal of Consciousness Studies} {\bf 15}(1), 17--38.

\smallskip
\noindent Misner C., Thorne K., and Wheeler J.A.~(1973): {\it Gravitation}, Freeman, New York. 

\smallskip
\noindent Mohrhoff U.~(1997): Interactionism, energy conservation, and the violation of physical laws. {\it Physics Essays} {\bf 10}, 651--665.

\smallskip
\noindent Nester J.M., So Lau Loi, and Vargas T.~(2008): On the energy of homogeneous cosmologies. {\it Physical Review D} {\bf 78}, 044035. 

\smallskip
\noindent Noether E.~(1918): Invariante Variationsprobleme. {\it Nachrichten der K\"oniglichen Gesellschaft der Wissenschaften zu G\"ottingen, Mathematisch-Physikalische Klas\-se}, pp.~235--257. Translated as ``Invariant Variation Problems'' by M.A.~Tavel, {\it Transport Theory and Statistical Physics} {\bf 1}, 183--207 (1971). 

\smallskip
\noindent Peebles P.J.E.~(1993): {\it Principles of Physical Cosmology}, Princeton University Press, Princeton.

\smallskip
\noindent Pitts J.B.~(2010): Gauge-invariant localization of infinitely many gravitational energies from all possible auxiliary structures. {\it General Relativity and Gravitation} {\bf 42}, 601--622. 

\smallskip
\noindent Pitts J.B.~(2011): Permanent underdetermination from approximate empirical equivalence in field theory: Massless and massive scalar gravity, neutrino, electromagnetic, Yang-Mills and gravitational theories. {\it British Journal for the Philosophy of Science} {\bf 62}, 259--299.

\smallskip
\noindent Pitts J.B.~(2016): Einstein's physical strategy, energy conservation, symmetries, and stability: ``But Grossmann \& I believed that the conservation laws were not satisfied''. {\it Studies in History and Philosophy of Modern Physics} {\bf 54}, 52--72. 

\smallskip
\noindent Pitts J.B.~(2019a): Conservation laws and the philosophy of mind: Opening the black box, finding a mirror. Forthcoming in {\it Philosophia}. 

\smallskip
\noindent Pitts J.B.~(2019b): Mental causation, conservation laws, and general relativity. Preprint.

\smallskip
\noindent Plantinga A.~(2007): Materialism and Christian belief. In {\it Persons: Human and Divine}, ed.~by  P.~van Inwagen and D.~Zimmerman, Oxford University Press, New York, pp.~99--141. (2007).

\smallskip
\noindent Rodrigues J.G.~(2014): There are no good objections to substance dualism. {\it Philosophy} {\bf  89}, 199--222.

\smallskip
\noindent Schmutzer E.~(1972): {\it Symmetrien und Erhaltungss\"atze der Physik}, Akademie-Verlag, Berlin. 

\smallskip
\noindent Schwartz J.M., Stapp H.P., and Beauregard M.~(2005): Quantum physics in neuroscience and psychology: A neurophysical model of mind-brain interaction. {\it Philosophical Transactions of the Royal Society B} {\bf 360}, 1309--1327. 

\smallskip
\noindent Searle J.R.~(2004): {\it Mind: A Brief Introduction}, Oxford University Press, Oxford.

\smallskip
\noindent Sorkin R.D.~(1988): Conserved quantities as action variations. In {\it Mathematics and General Relativity}, ed.~by J.A.~Isenberg, 
American Mathematical Society, Providence, pp.~23--37. 

\smallskip
\noindent Stephani H.~(1990): {\it General Relativity}, Cambridge University Press, Cambridge. 

\smallskip
\noindent Swinburne R.~(2019): The implausibility of the causal closure of the physical. {\it Organon} {\bf F 26}, 25--39.

\smallskip
\noindent von Wachter D.~(2006): Why the argument from causal closure against the existence of immaterial things is bad. In   {\it Science -- A Challenge to Philosophy?}, ed.~by H.J.~Koskinen, R.~Vilkko, and S.~Philstr\"om, Peter Lang, Frankfurt, pp.~113--124. 

\smallskip
\noindent Wald R.M.~(1984): {\it General Relativity}, University of Chicago Press, Chicago. 

\smallskip
\noindent Watkins E.~(1995): The development of physical influx in early eighteenth-century Germany:  Gottsched, Knutzen, and Crusius. {\it Review of Metaphysics} {\bf 49}, 295--339.

\smallskip
\noindent Watkins E.~(1998): From pre-established harmony to physical influx: Leibniz's reception in eighteenth century Germany. {\it Perspectives on Science} {\bf 6}, 136--203.

\smallskip
\noindent Westphal J.~(2016): {\it The Mind--Body Problem}, MIT Press, Cambridge.

\smallskip
\noindent Weyl H.~(1922): {\it Space--Time--Matter}, Methuen \& Company, London. Translated by H.L.~Brose. Reprinted by  Dover, New York (1952). 

\smallskip
\noindent Wigner E.~(1967): {\it Symmetries and Reflections}, Indiana University Press, Bloomington.

\smallskip
\noindent Wilson D.L.~(1999): Mind-brain interaction and violation of physical laws. {\it Journal of Consciousness Studies} {\bf 6}(8-9), 185--200.

\bigskip
\bigskip

\leftline{\sl Received: 18 May 2019}
\leftline{\sl Revised: 28 June 2019}
\leftline{\sl Accepted: 30 June 2019}
\smallskip
\leftline{\sl Reviewed by Jeffrey Koperski and another, anonymous, referee.}   
}

\newpage
\thispagestyle{empty}
\section*{}

\end{document}